\documentclass[aps,preprint]{revtex4}%
\usepackage{amsfonts}
\usepackage{amsmath}
\usepackage{amssymb}
\usepackage{graphicx}%
\providecommand{\U}[1]{\protect\rule{.1in}{.1in}}

\begin{document}
\title{Generating spacetimes from colliding sources}
\author{M. Halilsoy}
\email{mustafa.halilsoy@emu.edu.tr}
\author{V. Memari}
\email{vahideh.memari@emu.edu.tr}
\affiliation{Department of Physics, Faculty of Arts and Sciences, Eastern Mediterranean
University, Famagusta, North Cyprus via Mersin 10, Turkey}
\date{\today}

\begin{abstract}
Certain well-known spacetimes of general relativity (GR) are generated from
the collision of suitable null-sources coupled with gravitational waves. This
is a classical process underlying the full nonlinearity of GR that may be
considered alternative to the quantum creativity at a large scale.
Schwarzschild, de Sitter, anti de Sitter and the $\gamma$-metrics are given as examples.

\end{abstract}

\maketitle

\section{Introduction}

It is a general conviction in physics that particles are created by virtue of
quantum principles so that quantum fields are the basic creators of matter.
The prototype of the idea goes back to early quantum electrodynamics in which
a photon decays into a pair of electron-positron, and the opposite process of
annihilation. The argument can be extended to arbitrary bosons, massive and
massless. The spin content extends to cover the fermions also in this venture.
In all this argument it is the energy that plays the fundamental role so that
energy converts into mass according to E=mc%
${{}^2}$%
, the most famous equation on our planet. We should agree that quantum
mechanics (QM) and quantum field theory (QFT) are very successful in such
atomic dimensions. Whether this glory extends to cosmological, even our/human
scale, remains disputable. In the strong gravitational field regime, for
instance, the much-appraised Hawking process \cite{1} remains still to be
verified. Namely, near a classical black hole (BH) creation of a pair gives
rise to a temperature that is still to be observed. The analog BHs from field
theory ignore the extremely attractive force of a real BH. In other words, why
do quantum fields survive the destructive power of a BH? Can any other
fundamental force other than quantum play in the role of creation and
destruction of matter that acts also at intermediate scales? Our answer will
concentrate on the nonlinear interaction/collision of classical sources in
general relativity (GR) \cite{2}. The process of collision is so violent that
it creates a spacetime singularity to undermine the creative benefits of the
Planck scale quantum. In such an atmosphere what good a quantum field does
other than being crashed into the singularity.

In this paper, we mainly consider a class of metrics \cite{3}\cite{4} that
describe objects with arbitrary shapes. A spherically symmetric case is
naturally the simplest element corresponding to the well-known Schwarzschild
solution/BH. We introduce incoming fields moving at the speed of light,
colliding head-on to create the elements of $\gamma-$ metrics. The transition
from the space of colliding waves to the space of BHs was provided by the
Chandrasekhar-Xanthopoulos (CX) duality.\cite{5}. In accordance with the
spirit of GR, we elaborate on such a principle to derive basic spacetimes as a
result of colliding null sources coupled with gravitational waves. We provide
other examples such as de Sitter and anti de Sitter. Our entire process is
classical as the spacetimes are classical.

The organization of the paper is as follows. In section II we present the
examples of Schwarzschild and de Sitter spacetimes. Section III follows with
the cover of $\gamma-$metrics and we conclude the paper in section IV with a
Conclusion and discussion.

\section{The examples of Schwarzschild and de Sitter spacetimes}

\subsection{The Schwarzschild case}

We consider the line element%
\begin{equation}
ds^{2}=2\left(  1+\eta\right)  ^{2}dudv-\Delta\delta\left(  \frac{1+\eta
}{1-\eta}\right)  dx^{2}-\left(  \frac{1-\eta}{1+\eta}\right)  dy^{2}\text{,}
\label{1}%
\end{equation}
where $\eta=\sin\left(  u+v\right)  $, $\mu=\sin\left(  u-v\right)  $,
$\Delta=1-\eta^{2}$, $\delta=1-\mu^{2}$ and $\partial_{x}$, $\partial_{y}$ are
two spacelike Killing vectors. $\left(  u,v\right)  $ are the null coordinates
along which massless fields propagate. By letting $u\rightarrow u\Theta\left(
u\right)  $, $v\rightarrow v\Theta\left(  v\right)  $, where $\Theta\left(
z\right)  $ is the unit step function defined by%
\[
\Theta\left(  z\right)  =\left\{
\begin{array}
[c]{c}%
1\text{, \ \ \ \ \ \ \ \ \ }z>0\\
0\text{, \ \ \ \ \ \ \ \ \ }z<0
\end{array}
\right.
\]
\newline whose derivative is a Dirac delta function $\delta\left(  z\right)
$, we can formulate the problem of colliding waves. For this purpose we divide
spacetime into $4$ regions \cite{2} as follows:

\paragraph{Region I:}

$\left(  u<0\text{, }v<0\right)  $ is the flat, vacuum spacetime described by%
\begin{equation}
ds^{2}=2dudv-dx^{2}-dy^{2} \label{2}%
\end{equation}

\paragraph{Region II:}

$\left(  u\geq0\text{, }v<0\right)  $ is the incoming region with line element%
\begin{equation}
ds^{2}=2\left(  1+\eta\right)  ^{2}dudv-\left(  1-\eta\right)  \left(
1+\eta\right)  ^{3}dx^{2}-\left(  \frac{1-\eta}{1+\eta}\right)  dy^{2}\text{,}
\label{3}%
\end{equation}
where $\eta\left(  u\right)  =\sin\left(  u\Theta\left(  u\right)  \right)  $.
In an approprate null-tetrad of Newman-Penrose (NP) \cite{6} we have the only
nonzero Weyl component%
\begin{equation}
\psi_{4}\left(  u\right)  =-\delta\left(  u\right)  +\frac{3\Theta\left(
u\right)  }{\left(  1+\eta\right)  ^{5}} \label{4}%
\end{equation}
representing a gravitational wave composed of an impulsive $\left(
\delta\left(  u\right)  \right)  $ and shock $\left(  \Theta\left(  u\right)
\right)  $ components. All Ricci components vanish.

\paragraph{Region III:}

$\left(  u<0\text{, }v\geq0\right)  $ is the opposite incoming region, same as
II, with $u\rightarrow v$ and $\psi_{4}\rightarrow\psi_{0}$ so that the
non-zero Weyl component is%
\begin{equation}
\psi_{0}\left(  v\right)  =-\delta\left(  v\right)  +\frac{3\Theta\left(
v\right)  }{\left(  1+\eta\right)  ^{5}} \label{5}%
\end{equation}

\paragraph{Region IV:}

$\left(  u>0\text{, }v>0\right)  $ is the interaction region described by the
line element (\ref{1}).

By a coordinate transformation, we wish to show that the line element
(\ref{1}) is nothing but the Schwarzschild spacetime. We have first
\begin{equation}
4dudv=\frac{d\eta^{2}}{\Delta}-\frac{d\mu^{2}}{\delta} \label{6}%
\end{equation}
which yields $\eta=\sin\left(  u+v\right)  $ and $\mu=\sin\left(  u-v\right)
$.

Next, we apply the transformation%
\begin{align}
1+\eta &  =\frac{r}{m}\nonumber\\
\sin\left(  u-v\right)   &  =\cos\theta\nonumber\\
x &  =\frac{\varphi}{\sqrt{2}}\nonumber\\
y &  =\frac{t}{m\sqrt{2}}\label{7}%
\end{align}
Here, $m$ emerges as a scaling constant that will be identified as the mass in
the resulting metric which is recognized as the Schwarzschild spacetime.%
\begin{equation}
2m^{2}ds^{2}=\left(  1-\frac{2m}{r}\right)  dt^{2}-\frac{dr^{2}}{\left(
1-\frac{2m}{r}\right)  }-r^{2}\left(  d\theta^{2}+\sin^{2}\theta d\varphi
^{2}\right)  \label{8}%
\end{equation}
To summarize, the collision of $\psi_{4}\left(  u\right)  $ in (\ref{4}) and
$\psi_{0}\left(  v\right)  $ in (\ref{5}) creates a region that is
transformable to the Schwarzschild black hole. Admittedly our technological
capacity for the time being is poor enough to provide this much strong
gravitational waves, collide them, and create a Schwarzschild black hole on
Earth. But this does \ not exclude the possibility of such a process in the
cosmos or colliding other sources to make new spacetimes.

\subsection{The de Sitter cosmology}

As a supplementary example we consider the collision problem of two null
shells \cite{7}\cite{8}\cite{9} described by the following incoming metrics

\subsubsection{Region II:}%

\begin{equation}
ds^{2}=\frac{1}{\left(  1+\alpha u\Theta\left(  u\right)  \right)  ^{2}%
}\left(  2dudv-dx^{2}-dy^{2}\right)  \text{,} \label{9}%
\end{equation}
$\ $

where $\alpha=$constant., with the non-vanishing Ricci component%
\begin{equation}
\phi_{22}\left(  u\right)  =\alpha\delta\left(  u\right)  \label{10}%
\end{equation}

\subsubsection{Region III:}%

\begin{equation}
ds^{2}=\frac{1}{\left(  1+\beta v\Theta\left(  v\right)  \right)  ^{2}}\left(
2dudv-dx^{2}-dy^{2}\right)  \text{,} \label{11}%
\end{equation}

where $\beta=$constant., with the Ricci term%
\begin{equation}
\phi_{00}\left(  v\right)  =\beta\delta\left(  v\right)  \label{12}%
\end{equation}
It is not difficult now to choose $\alpha=\beta$, and show that the
interaction region line element takes the following form

\subsubsection{Rigion IV:}%

\begin{equation}
ds^{2}=\frac{1}{\phi^{2}}\left(  2dudv-dx^{2}-dy^{2}\right)  \text{,}
\label{13}%
\end{equation}

where $\phi=1+\alpha\left(  u\Theta\left(  u\right)  +v\Theta\left(  v\right)
\right)  $ with the non-vanishing NP scalors%
\begin{align}
\phi_{22} &  =\frac{\alpha\delta\left(  u\right)  }{1+\alpha v}\nonumber\\
\phi_{00} &  =\frac{\alpha\delta\left(  v\right)  }{1+\alpha u}\nonumber\\
\Lambda &  =\frac{R}{24}=\frac{\alpha^{2}}{\phi^{2}}\label{14}%
\end{align}
By applying the transformation%
\begin{align}
1+\alpha\left(  u+v\right)   &  =e^{\lambda t}\nonumber\\
\alpha\left(  u-v\right)   &  =\lambda z\label{15}%
\end{align}
where $\lambda$ =$\sqrt{2}\alpha=$constant., we obtain the de Sitter spacetime%
\begin{equation}
ds^{2}=dt^{2}-e^{-2\lambda t}\left(  dx^{2}+dy^{2}+dz^{2}\right)  \label{16}%
\end{equation}
Closely related to the de Sitter (dS) we have the anti de Sitter (AdS) which
emerges as a result of a colliding null-shells with partly reflection. The
spacetime globally is described in the form%
\begin{equation}
ds^{2}=\frac{1}{\left(  1+\alpha\left(  u\Theta\left(  u\right)  -v\beta
\Theta\left(  -v\right)  \right)  \right)  ^{2}}\left(  2dudv-dx^{2}%
-dy^{2}\right)  \label{17}%
\end{equation}
The non-zero NP quantities are given as follows%
\begin{align}
\Lambda &  =\frac{R}{24}=-\alpha^{2}\Theta\left(  u\right)  \Theta\left(
-v\right)  \nonumber\\
\phi_{22} &  =\alpha\left(  1-\alpha v\Theta\left(  -v\right)  \right)
\delta\left(  u\right)  \nonumber\\
\phi_{00} &  =\alpha\left(  1+\alpha u\Theta\left(  u\right)  \right)
\delta\left(  v\right)  \label{18}%
\end{align}
It is observed that besides the criss-crossing null-shells, the Region II,
$\left(  u\geq0\text{, }v<0\right)  $ possesses a constant scalar curvature
which can be identified as the AdS spacetime. This is obtained easily by the
transformation.%
\begin{align}
1+\alpha\left(  u-v\right)   &  =z\nonumber\\
\alpha\left(  u+v\right)   &  =t\nonumber\\
x &  \rightarrow\frac{1}{\sqrt{2}\alpha}x\nonumber\\
y &  \rightarrow\frac{1}{\sqrt{2}\alpha}y\label{19}%
\end{align}
which gives the manifestly conformal flat form of the AdS as
\begin{equation}
ds^{2}=\frac{1}{2\alpha^{2}z^{2}}\left(  dt^{2}-dx^{2}-dy^{2}-dz^{2}\right)
\label{20}%
\end{equation}

\section{Emergent $\gamma-$metric from colliding fields}

We go now one step forward and consider the non-spherically symmetric
$\gamma-$metric as a yield of two colliding fields consisting of gravitational
waves coupled with null sources. To formulate this case as an initial value
problem of collision we consider the vacuum metric%
\begin{equation}
ds^{2}=\sqrt{\Delta}e^{N}\left(  \frac{d\eta^{2}}{\Delta}-\frac{d\mu^{2}%
}{\delta}\right)  -\sqrt{\Delta\delta}\left(  \chi dx^{2}+\chi^{-1}%
dy^{2}\right)  \label{21}%
\end{equation}
in which the metric functions are (following the notation of (\ref{1}))%
\begin{align}
\chi &  =\sqrt{\Delta\delta}\left(  \frac{1+\eta}{1-\eta}\right)  ^{\gamma
}\nonumber\\
e^{N} &  =\left(  \frac{1+\eta}{1-\eta}\right)  ^{\gamma}\Delta^{\gamma
^{2}-\frac{1}{2}}\mid\Delta-\delta\mid^{1-\gamma^{2}}\label{22}%
\end{align}
Here $\gamma=$costant. $\left(  0<\gamma<\infty\right)  $ is the distortion
parameter and as before, $\Delta=1-\eta^{2}$ and $\delta=1-\mu^{2}$. With an
appropriate transformation, this can be expressed in the standard form of the
Schwarszchild coordinates \cite{10}. Our task now is to determine the
colliding incoming waves/sources that make this line element. We define the
null coordinates $\left(  u,v\right)  $ through%
\begin{align}
2du &  =\frac{d\eta}{\sqrt{\Delta}}-\frac{d\mu}{\sqrt{\delta}}\nonumber\\
2dv &  =\frac{d\eta}{\sqrt{\Delta}}+\frac{d\mu}{\sqrt{\delta}}\label{23}%
\end{align}
and upon integration we obtain $\eta\left(  u,v\right)  $ / $\mu\left(
u,v\right)  $ as
\begin{align}
\eta &  =\sin\left(  u+v+c_{1}\right)  \nonumber\\
\mu &  =\sin\left(  -u+v+c_{2}\right)  \label{24}%
\end{align}
where $\left(  c_{1},c_{2}\right)  $ are constants related to the incoming
null sources. It is shown that for $\gamma\neq1$, which amounts to the case of
oblate $\left(  \gamma>1\right)  $ or prolate $\left(  \gamma<1\right)  $
objects we must have $0\neq c_{1}\neq c_{2}\neq0$. The spherical symmetry is
recovered for $\gamma=1$ with $c_{1}=c_{2}=0$. After some trivial scaling of
coordinates, the line element (\ref{21}) can be expressed as%
\begin{equation}
ds^{2}=2e^{-M}dudv-A^{2}\left(  u,v\right)  dx^{2}-B^{2}\left(  u,v\right)
dy^{2}\text{,}\label{25}%
\end{equation}
where%
\begin{align}
e^{-M} &  =\sqrt{\Delta}e^{N}\nonumber\\
A^{2} &  =\sqrt{\Delta\delta}\chi\nonumber\\
B^{2} &  =\sqrt{\Delta\delta}\chi^{-1}\label{26}%
\end{align}
The incoming fields of Region II can be determined by choosing a null tetrad
of the form,%
\begin{align}
l_{\mu}  & =e^{-M\left(  u\right)  }\delta_{\mu}^{u}\nonumber\\
n_{\mu}  & =\delta_{\mu}^{v}\nonumber\\
m_{\mu}  & =\frac{1}{\sqrt{2}}\left(  A\left(  u\right)  \delta_{\mu}%
^{x}+iB\left(  u\right)  \delta_{\mu}^{y}\right)  \nonumber\\
\overline{m}_{\mu}  & =\frac{1}{\sqrt{2}}\left(  A\left(  u\right)
\delta_{\mu}^{x}-iB\left(  u\right)  \delta_{\mu}^{y}\right)  \label{27}%
\end{align}
and by interchanging $u\leftrightarrow v$ we obtain the non-zero quantities
for the Region III. The non-vanishing spin coefficients in Region II are
$\left(  \lambda,\mu\right)  $ given by%
\begin{align}
\mu & =\frac{1}{2}\Theta\left(  u\right)  e^{M}\left[  -\tan\left(
u\Theta\left(  u\right)  +c_{1}\right)  +\tan\left(  -u\Theta\left(  u\right)
+c_{2}\right)  \right]  \nonumber\\
\lambda & =\mu+\frac{\gamma\Theta\left(  u\right)  e^{M}}{\cos\left(
u\Theta\left(  u\right)  +c_{1}\right)  }\label{28}%
\end{align}
The non-zero Weyl and Ricci components are%
\begin{align}
\psi_{4}\left(  u\right)    & =-\lambda_{u}e^{M}-2\lambda\mu\nonumber\\
\phi_{22}\left(  u\right)    & =-\mu_{u}e^{M}-\lambda^{2}-\mu^{2}\label{29}%
\end{align}
in which $\lambda_{u}=\frac{\partial\lambda}{\partial u}$ and $\mu_{u}%
=\frac{\partial\mu}{\partial u}$. The details of $\psi_{4}\left(  u\right)  $
does not interest us here which briefly consists of superposition of an
impulsive and shock gravitational wave.

In other words as in the Schwarzschild case $\psi_{4}\left(  u\right)  $
consists of impulse and shock components with much more complicated
coefficients. The exact form of $\phi_{22}\left(  u\right)  $ is given by%
\begin{equation}
\phi_{22}\left(  u\right)  =\frac{1}{2}e^{2M\left(  0\right)  }\delta\left(
u\right)  \left(  \tan c_{1}-\tan c_{2}\right)  \text{,}\label{30}%
\end{equation}
where
\begin{equation}
e^{-M\left(  0\right)  }=\left(  \frac{1+\sin c_{1}}{1-\sin c_{1}}\right)
^{\gamma}\left(  \sin^{2}c_{2}-\sin^{2}c_{1}\right)  ^{1-\gamma^{2}}\left(
\cos c_{1}\right)  ^{2\gamma^{2}}\text{,}\label{31}%
\end{equation}
which represents a physically acceptable source for $\tan c_{1}>\tan c_{2}$.
Upon collision of Region II fields ($\psi_{4}\left(  u\right)  $ plus
$\phi_{22}\left(  u\right)  $) with those of Region III ($\psi_{0}\left(
v\right)  $ plus $\phi_{00}\left(  v\right)  $) we obtain the Region IV metric
given in (\ref{21}). Note that for ( $u>0$, $v>0$) we ignore the impulsive
terms that are effective on the null boundaries. The transformation (\ref{7})
can also be used here to transform (\ref{25}) into (\ref{21}) in the
Schwarzschild coordinates, which is the $\gamma-$metric.

\section{Conclusion and discussion}

Can we generate all meaningful spacetime metrics of GR by the procedure of
colliding classical sources?. Recalling the famous remark by Penrose \cite{11}
that every spacetime admits a plane wave limit, encourages us to ensure the
answer in the affirmative. That is, each Penrose limit for $v<0$ / $u<0$ can
be considered as the incoming wave to formulate the problem as a collision.
Each case, however, must be studied separately to find the necessary
incoming/colliding sources, as in the case of colliding particles in
high-energy physics. We remark that even the Kerr metric is not exception to
our analysis \cite{5}\cite{12}. We recall that particles in Quantum Theory are
probability fields whereas in GR they are gravitational fields. Our emphasis
in this study is to advocate the view that without reference to quantum, we
can generate a spacetime such that the mass/charge in it emerges through
scaling of coordinates. The drawback of our model is that to obtain an exact
solution we assume certain symmetry which makes the mutual focusing so strong
that yields also spacetime singularities. Extension of our model to string
theory collisions which excludes point charges may eliminate the
singularities. Finally, the inverse problem of collision may be argued; an
instability through perturbation or by an intruding agent into the spacetime
causes decays into its components or collapse into a spacetime singularity.

\end{document}